\documentclass[aps,pra,twocolumn,preprintnumbers,amsmath,amssymb,showpacs]{revtex4}
\usepackage{graphicx}
\usepackage{float}
\usepackage{pstricks}           
\usepackage{dcolumn}            
\usepackage{bm}                 
\usepackage{hyperref}           
\usepackage[latin1]{inputenc}   
\usepackage[T1]{fontenc}
\usepackage[british]{babel}
\usepackage{soul}
\usepackage{rotating}
\usepackage{subfigure}
\usepackage{slashed}
\usepackage{color} 

\begin{document}

\title{Intensity Effect of Laser-assisted Double Ionization of Helium Atoms by Ultrashort XUV Pulses}
\author{ Fengzheng Zhu$^{1,2}$, Genliang Li$^{1}$, Aihua Liu$^{1}$\thanks{Corresponding author. E-mail:~aihualiu@jlu.edu.cn}}
\affiliation{$^{1}${Institute of Atomic and Molecular Physics, Jilin University, Changchun 130012, China}\\  
$^{2}${School of Mathematics and Physics, Hubei Polytechnic University, Huangshi Hubei 435003, China } }
\date{\today}

\begin{abstract}
We investigate the laser-assisted single XUV photon double ionization of ground state helium. By solving full six-dimensional time-dependent schr{\"o}dinger equation, we study the correlation and angular distribution. The discussion in joint energy and angular distributions reveals the competition in odd and even parity double ionization processes in the presence of weak laser field. The emission angle  between two photoelectrons can be adjusted by the laser parameters.  We depicts how the laser field enhance and/or enable the back-to-back and side-by-side emission of double ionization.

\end{abstract}

\maketitle
\section{Introduction}
Investigations of double ionization (DI) of helium atom with extreme ultraviolet (XUV) laser field~\cite{Bengtsson2019} have been launched both experimentally~\cite{Huillier1982,Schwarzkopf1993a,feuerstein2001}  and theoretically~\cite{Maulbetsch_1993,McCurdy2004,Avaldi_2005} for more than two decades.
Recently, the  experimental advances in generation of  XUV light from free-electron lasers technology ~\cite{emma2010first,Rosenblum2015,Shintake2008b} and the application of the high-order harmonic generation~\cite{Lewenstein1994}  provide opportunities to observe single- and multiple ionization of atoms and molecules experimentally~\cite{Briggs1999,Schwarzkopf1993a,McPherson1987b}.
Hasegawa~\cite{Hasegawa2005} firstly performed an experiment to measure the two-photon DI cross sections of helium
 exposed to XUV pulses with a photon energy of 42 eV.
In particular, major theoretical researches have been focused on calculations of the triply differential cross section (TDCS) for the DI of helium, for example, TDCSs calculated by Huetz \textit{et al.}~\cite{AHuetz1991} based on the Wannier theory~\cite{Wannier1953}, TDCSs obtained from time-dependent close-coupling simulations by Palacios \textit{et al.}~\cite{Palacios2008} and the convergent close-coupling calculations by Kheifets and Bray~\cite{Braeuning1998bbb}. All these works were found to be in good agreement with the angular distribution in the absolute TDCSs measured by Br\"{a}uning \textit{et al.}~\cite{Braeuning1998bbb}.


Recent progress in experimental physics has performed pump-probe scheme with the combined infrared (IR) and extremely ultraviolett (XUV) laser fields for investigating electron dynamics in ultrashort time scales, such as above threshold ionization(ATI)~\cite{Meyer2008bbb,Becker2002,Pi2014}, streaking camera \cite{Jason2018}, high-order harmonic generation~\cite{kim2005highlybbb,Wang2020}, and nonsequential double ionization(NSDI)~\cite{Liao2012,Zhang2014} as well. The enhancement of emission of fast photoelectrons and sensitive influences on the photoelectron energy distribution caused by the additional IR laser field was reported by Hu~\cite{Hu2013}. Also, the IR pulse's promotion of side-by-side and back-to-back  emission is illustrated by using of the FE-DVR method for numerically solving the TDSE equation in full dimensionality~\cite{rescigno2000numerical,feist2008nonsequential,pazourek2011universal}. In addition, the selection rules~\cite{Maulbetsch_1993,Maulbetsch_1994}, comprehensive numerical studies as well as the joint angular distributions (JADs) with different energy sharing of the emitted electrons are learned~\cite{Zhang2011}.
In previous work~\cite{Liu2014}, liu \textit{et al.} studied the DI process of an helium in IR-assisted XUV laser field by using of the finite-element(FE) discrete-variable-representation(DVR) method for numerically solving the time-dependent Schr\"oinger equation(TDSE) in full dimensionality.
It has been found that the assistant IR pulse  promotes the side-by-side and enables back-to-back emission. Also, we analyzed the dependence of JADs and mutual photoelectron angular distributions(MADs) on the energy sharing of the emitted electrons. However, the main purpose of the present paper is to investigate how the weak IR field change the energy distribution and emission pattern of photoelectrons.More recently,
Jin  \textit{et al.}~\cite{Jin2018} investigated the role of IR and XUV laser field respectively,  in which case the NSDI process is described as an ATI followed by a laser-assisted collision.

An emission geometry for co-planar emission has been developed in Ref.~\cite{Liu2014}, which
distinguishes for typical electron emission types as back-to-back, side-by-side, conic and symmetric emission, and supposes the electron is ejected with angles $\theta_{i}$,  relative to the polarization directions of XVU and IR pulses.
Investigations of these emission patterns allow one to investigate  the dynamics in two-photon DI process more precisely. The electron emission type in sequential and nonsequential  two-photon DI process has been discussed by calculating the JAD at equal energy sharing.
Jin~\cite{Jin2016} showed the contributions of forward and backward collisions to the NSDI probability  in the side-by-side and back-to-back patterns.

Here, we carry calculations of probability density in joint energy distribution(JED) and JAD for the photoelectrons ejected during the ionization of helium atoms by an intense XUV pulse in the presence of a weak IR laser pulse, and we will discuss how the emission pattern varies with the increasing intensity of the assisting IR laser field. Also, we study the dependency of JADs and MADs on the energy sharing between the photoelectrons, and find the enhancement of back-to-back emission pattern during the DI process at extremely unequal energy sharing  is attributed to the ejected electrons with odd-parity.

The rest of this paper is organized as follows:
Theoretical methods for the numerical solution of the full-dimensional TDSE for the two emitted electrons during the ionization are presented in section 2. The results for the energy- and angle-differential correlated electron-emission probabilities for DI process due to laser-assisted XUV pulses are shown and discussed in section 3. Conclusions are drawn in section 4.

\section{Theory and numerical implementation}\label{sec:Theo}
The motion of two-electron atoms driven by the laser-assisted XUV laser field is described by the TDSE in atomic units
\begin{equation}
i\frac{\partial \Phi(\textbf{r}_1,\textbf{r}_2;t)}{\partial t} =
H\Phi(\textbf{r}_1,\textbf{r}_2;t)
\end{equation}
The Hamiltonian of the system can be separated into two contributions
\begin{equation}
H = H_{A} + V_{I}.
\end{equation}
The first term describes the undisturbed atomic system,
\begin{equation}
H_{A} = -\nabla_{1}^{2}/2-Z/r_{1}-\nabla_{2}^{2}/2-Z/r_{2}+1/|\textbf{r}_1-\textbf{r}_2|
\end{equation}
with $Z=2$ being the nuclear charge of He.
The second term $V_{I}$ which describing the time-dependent atom-electric coupling in the dipole length-gauge is expressed as:
\begin{equation}
V_I=-(\textbf{E}_{\text{XUV}}(t)+\textbf{E}_{\text{IR}}(t))\cdot(\textbf{r}_1+\textbf{r}_2).
\end{equation}
Here the XUV and IR pulses are both assumed to be linearly polarized with sine-squared temporal profiles and given by£º
\begin{equation}
\textbf{E}_{\alpha}(t)= \begin{cases}
{E}_{0\alpha}\sin^2(\frac{\pi t}{\tau_\alpha})\cos(\omega_a t+\varphi_a), & \mbox{if } 0<t<\tau_a \\
        0, & \mbox{else,} \end{cases}
        \label{eq:pulse}
\end{equation}
with ${E}_{0\alpha}$,$\tau_\alpha$, $\varphi_\alpha$ and $\omega_\alpha$ ($\alpha$ = IR, XUV) being the electric-filed amplitude, pulse lengths, carrier-envelope phases, and frequencies of the XUV and IR field, respectively.

 For the two-electron system exposed to the XUV and IR laser field, the wave function was expanded in terms of the bipolar spherical harmonic \cite{Hu2010,Zhang2011,Liu2014}
 \begin{equation}
\Phi(\textbf{r}_1,\textbf{r}_2;t)=
\sum_{LM}\sum_{l_1,l_2}\frac{\psi_{l_1l_2}^{(LM)}(r_1,r_2;t)}{r_1r_2}
\mathcal{Y}_{l_1l_2}^{L,M}(\hat{\textbf{r}}_1,\hat{\textbf{r}}_2)
\label{eq:PHI}
\end{equation}
where $\mathcal {Y}_{l_1l_2}^{LM}(\hat{\textbf{r}}_1,\hat{\textbf{r}}_2)=\sum_{m_1m_2}C_{l_1m_1l_2m_2}^{LM}Y_{l_1m_1}(\hat{\textbf{r}}_1)
Y_{l_2m_2}(\hat{\textbf{r}}_2)$ is the coupling of the two electrons' individual angular momenta. In above $l_{i}$ and $m_{i}(i=1,2)$ are the quantum numbers, $Y_{l_im_i}(\hat{\textbf{r}}_i)$ is the ordinary spherical harmonics and $C_{l_1m_1l_2m_2}^{LM}$ is the Clebsch-Gordan coefficient.
In the case of initial singlet state of helium, the quantum number $M=m_{1}+m_{2}$ is conserved to be $0$ for  linearly polarized laser field. Moreover,  due to the spatial symmetry of wave function originates from the indistinguishability between the two Fermions, the sum over $L$ is restricted to even values of $L-l_1-l_2$. The initial singlet-spin state is obtained by replacing the real time $t$ in TDSE with the imaginary time $\tau=it$ \cite{Liu2014}.
The numerical solution of the final wave function can be accurately obtained by application of FEDVR \cite{rescigno2000numerical,Hu2010,Zhang2011,Liu2014,Liu2015,Zhou2019}.

Then the DI probability for double ionization corresponding to final state with momenta $\textbf{k}_1,\textbf{k}_2$ can be given by projecting the final wave function to the asymptotic two-electron wave function for a long time after the termination of the pulses.
For the purpose of removing spurious contribution caused by nonorthogonality of the approximate asymptotic wave function and the initial state, we rewrite the final wave function  with  exclusion of the overlap between the initial state and the final state.

Finally, we evaluate the correlated energy distribution by integrating over all angles:
\begin{equation}
P(E_1,E_2) = \frac{1}{k_1k_2}\int d\Omega_1d\Omega_2
P(\textbf{k}_1,\textbf{k}_2), \label{eq:edist}
\end{equation}
where $E_1~=~k_1^2/2$ and $E_2~=~k_2^2/2$ are the final (asymptotic)
energies of the two emitted electrons.

\section{Results and Discussion}\label{sec:Res}
Our previous studies have focused on the modification of the energy and angular distribution of photoelectrons caused by a weak assisted-IR laser field during the DI of helium in an intense XUV pulse. It is worth pointing out that the certain effective number of absorbed minus emitted IR photon can be represented by each stripe in the sideband pattern of the joint energy distribution.


We analyze the dependence of joint and mutual photoelectron angular distributions on the intensity of the assisted-IR laser field for the DI of helium atoms by short XUV pulses in co-planar emission geometry.
For the sake of simplicity, we suppose that the XUV pulse and IR pulse have overlapping cosine-squared temporal profiles with identical phases $\varphi_{XUV}=\varphi_{IR}=0^{\circ}$.
We use the IR and XUV pulses with pulse length as 2.6fs and 0.46fs, respectively.
The peak intensity of the XUV laser field is chosen at $10^{13}$ W/cm$^{2}$,
And the central photon energy of the XUV laser and IR field is supposed to be $\hbar\omega_{XUV}=89$ eV and $\hbar\omega_{IR}=1.61$ eV, respectively.


\begin{figure}[H]
\begin{center}
\includegraphics[width=\columnwidth]{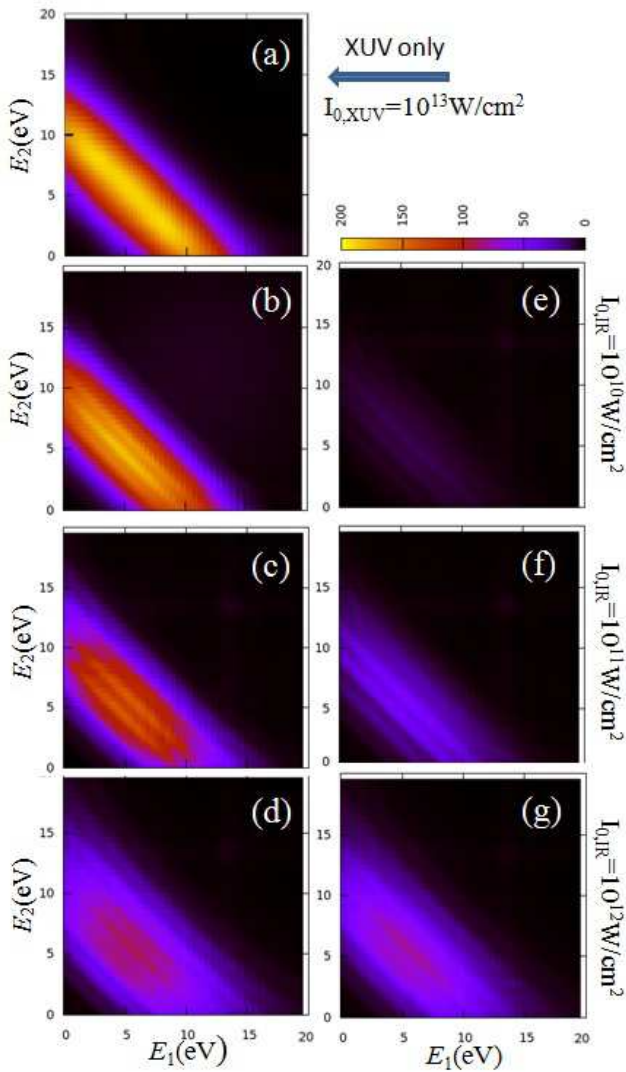}\\[4pt]  
\end{center}
\caption{Joint energy distribution of the two emitted electrons ionized by the XUV laser field only is presented in (a) with peak intensity $10^{13}$ W/cm$^{2}$. Comparisons of joint energy distribution for odd-parity electrons (b,c,d) and even-parity electrons (e,f,g) in different assited-IR laser field intensities  $10^{10}$ W/cm$^{2}$(b,e), $10^{11}$ W/cm$^{2}$(c,f) and $10^{12}$ W/cm$^{2}$(d,g), where the peak intensity of the XUV pulses is $10^{13}$ W/cm$^{2}$,  and the photon energy of the XUV laser field is 89 eV.
}\label{fig:jed01}
\end{figure}


We first consider the JEDs of the photoelectrons driven by an XUV pulse alone (i.e. without the assisted-IR field) of central energy 89 eV in Figs.~\ref{fig:jed01} (a), in which the largest DI yields are uniformly located at $E_{1}+E_{2}=10$ eV signifying a nonsequential DI.
When the assisted-IR filed is turned on, the DI yields turn to be composed by sidebands which are located at $E_{1}+E_{2}=11$ eV+$N\hbar\omega_{IR}$ corresponding to the absorption of (effectively) one XUV photon and N IR photons.
These sideband patterns result from different pathway to implement the DI with a certain number $N_{IR}$  of absorbed minus emitted IR photons.
However, these stripes are not easily resolved because the energy of the assisted-IR photon $\hbar\omega_{IR}=1.61$ eV is much less than the FWHM of the XUV pulse.

To display this sideband, the decompositions of JEDs into contributions with even or odd numbers of absorbed minus emitted IR photons are showed separately at different laser intensity [see Figs.~\ref{fig:jed01}(b-g)].  The left panels are JEDs for odd parity, and the right panels show JEDs for even parity. From top to bottom, the IR intensities are $0$, $10^{10}$, $10^{11}$ and $10^{12} $W/cm$^{2}$, respectively.
In each graph of Figs.~\ref{fig:jed01}(b-g), the separation between the neighboring stripes equal to the energy of two IR photons (3.2 eV).
One can first find that, with increasing IR-lase intensity, the odd-parity probability density in the JEDs decreases. By contrast, for the photoelectron absorbing even numbers of XUV photons P($E_1$,$E_2$) displays a decreasing trend. This means that the weak assisting IR-laser field does not change the DI yield obviously, but affects the parity of photoelectron and changes their distribution.
In addition, the comparison between the two panels demonstrate that the odd-parity and even-parity photoelectrons are competing. In addition, stronger IR fields can boost the even-parity photoelectron. Therefore, at the intensity $10^{12}$ W/cm$^{2}$, the odd-parity and even-parity probability densities are comparable in magnitude.
However, if we keep increasing the IR-laser intensity, e.g. to $10^{13}$ W/cm$^{2}$, the DI-probability increases due to the strong IR-laser field can be significant, and many-photon process will be important or even dominant. Since we have to employ many more partial waves to describe such a process, no numerical results for stronger assisted IR field are shown in this paper. And the numerical effort being significant and requiring access to computational resource that are not locally available to us.

\begin{figure}[H]
\begin{center}
\includegraphics[width=\columnwidth]{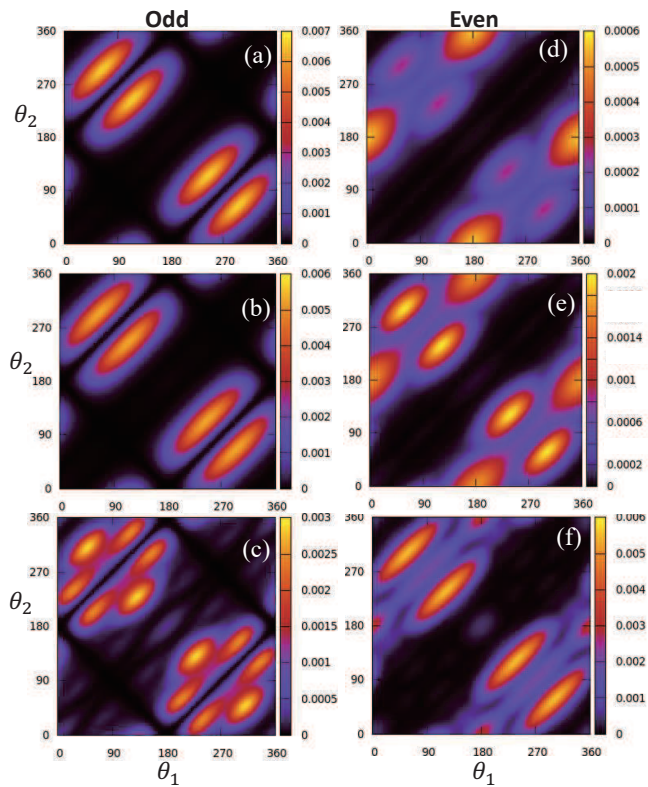}
\end{center}
\caption{Joint angular distributions of two photoelectrons at equal energy sharing with odd-parity (a-c) and even parity (d-f) for different assisted-IR laser intensities $I_{IR}=10^{10}$ W/cm$^{2}$ (a,d), $I_{IR}=10^{11}$ W/cm$^{2}$ (b,e), and  $I_{IR}=10^{12}$ W/cm$^{2}$ (c,f), where the peak intensity of the XUV laser field is $10^{13}$ W/cm$^{2}$.}\label{fig:JAD01}
\end{figure}

Then the JAD were calculated for the DI of helium using a XUV laser pulse at 89 eV and an intensity of $10^{13}$ W/cm$^{2}$ in the presence of an additional IR laser field with equal energy sharing.
The results are displayed with odd- and even-parity separately for peak intensity of $I_{IR}=10^{10}$ W/cm$^{2}$ [Figs.~\ref{fig:JAD01} (a,d)], $I_{IR}=10^{11} $ W/cm$^{2}$ [Figs.~\ref{fig:JAD01} (b,e)], and  $I_{IR}=10^{12}$ W/cm$^{2}$ [Figs.~\ref{fig:JAD01} (c,f)].
According to the pioneering work by Huetz {\it et al.}, the structure of the photoelectron angular distributions for single-photo double ionization was not only constrained to the selection rules, but also results from the combined action of symmetrical and antisymmetrical components with respect to electron exchange.

As illustrated in Figs.~\ref{fig:JAD01}(a-c) and Figs.~\ref{fig:JAD01}(d-f),  the increased peak intensity of the assisted-IR laser field alter the structure of photoelectron angular distributions.
The most significant change is that with increasing IR peak intensity the values of the peaks for the odd parity decreases, while the values for the even parity increase.
Intuitively, we may interpret these results as implying that the DI probability of the odd-parity electron is gradually suppressed by the even-parity electron with stronger IR field.
However, the total double ionization probability is almost unchanged.  That means the DI probability is independent on the intensity of the IR laser field.
This is consistent with the results have been shown in the JEDs discussed above.

The second change is that the positions of the main peaks change with peak density of the IR laser field.
For the case of odd-parity, the four main peaks are distributed along the line of  $\theta_{1}+\theta_{2}=360^{\circ}$, and the separation between the main peaks becomes larger as the stronger IR field increases.  Particularly  when the intensity of the IR laser field is taken as $10^{12}$ W/cm$^{2}$, the symmetric emission is fading to compete with back-to-back emission.

The figures for the even-parity electrons changes drastically as intensity of the IR laser field increase.
We can see that the dominant emission pattern changes gradually from the `back-to-back' (as shown in Fig.~\ref{fig:JAD01}(d)) to the symmetric pattern (as shown in Fig.~\ref{fig:JAD01}(f))
This change implies that, although DI produced for the odd-parity electron is competed with that for the even-parity electron, the two-photoelectron emission is dominant in the symmetric pattern.

\begin{figure}[H]
\begin{minipage}[c]{0.5\linewidth}
\centering
\includegraphics[width=.99\columnwidth]{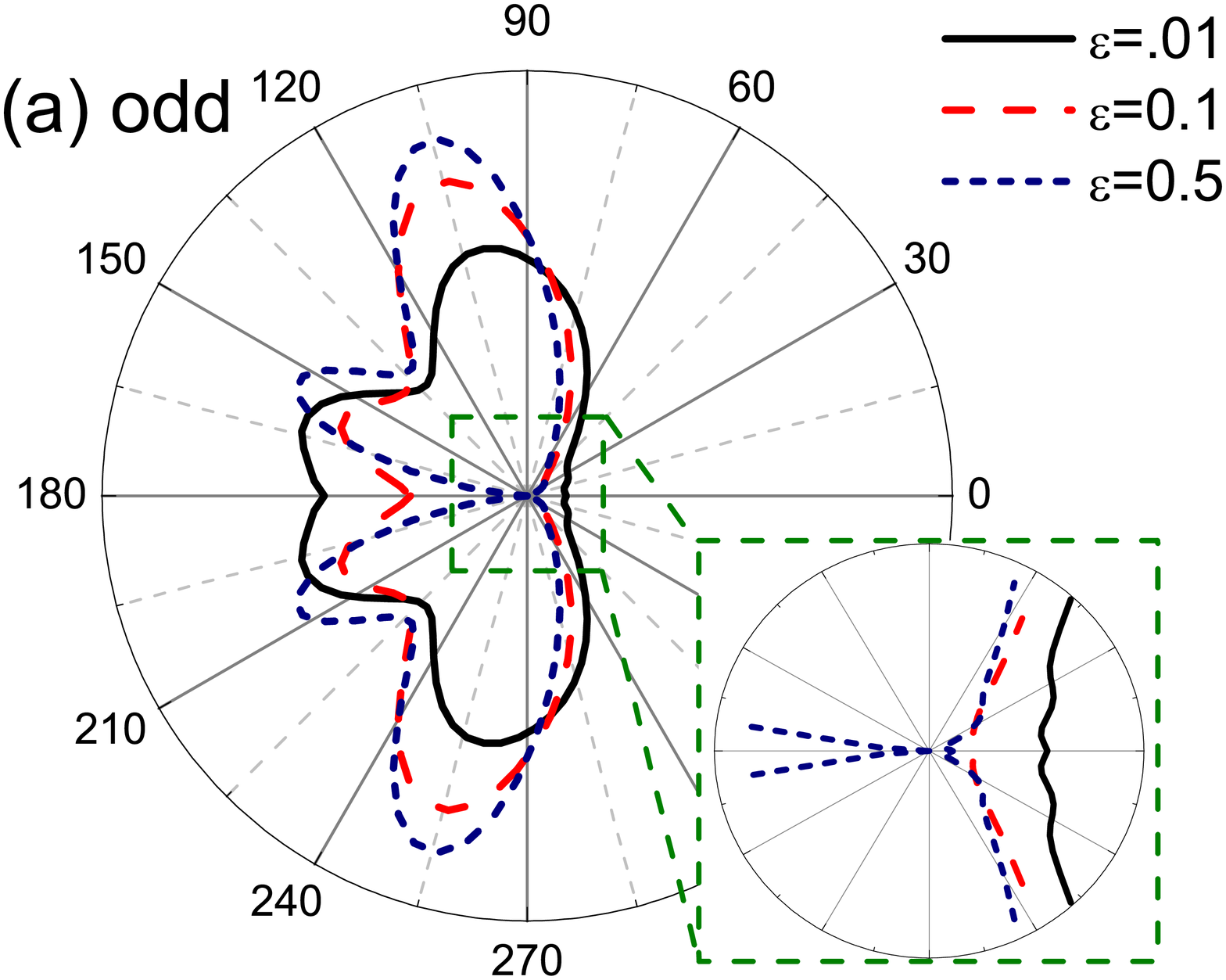}
\end{minipage}%
\begin{minipage}[c]{0.5\linewidth}
\centering
\includegraphics[width=.99\columnwidth]{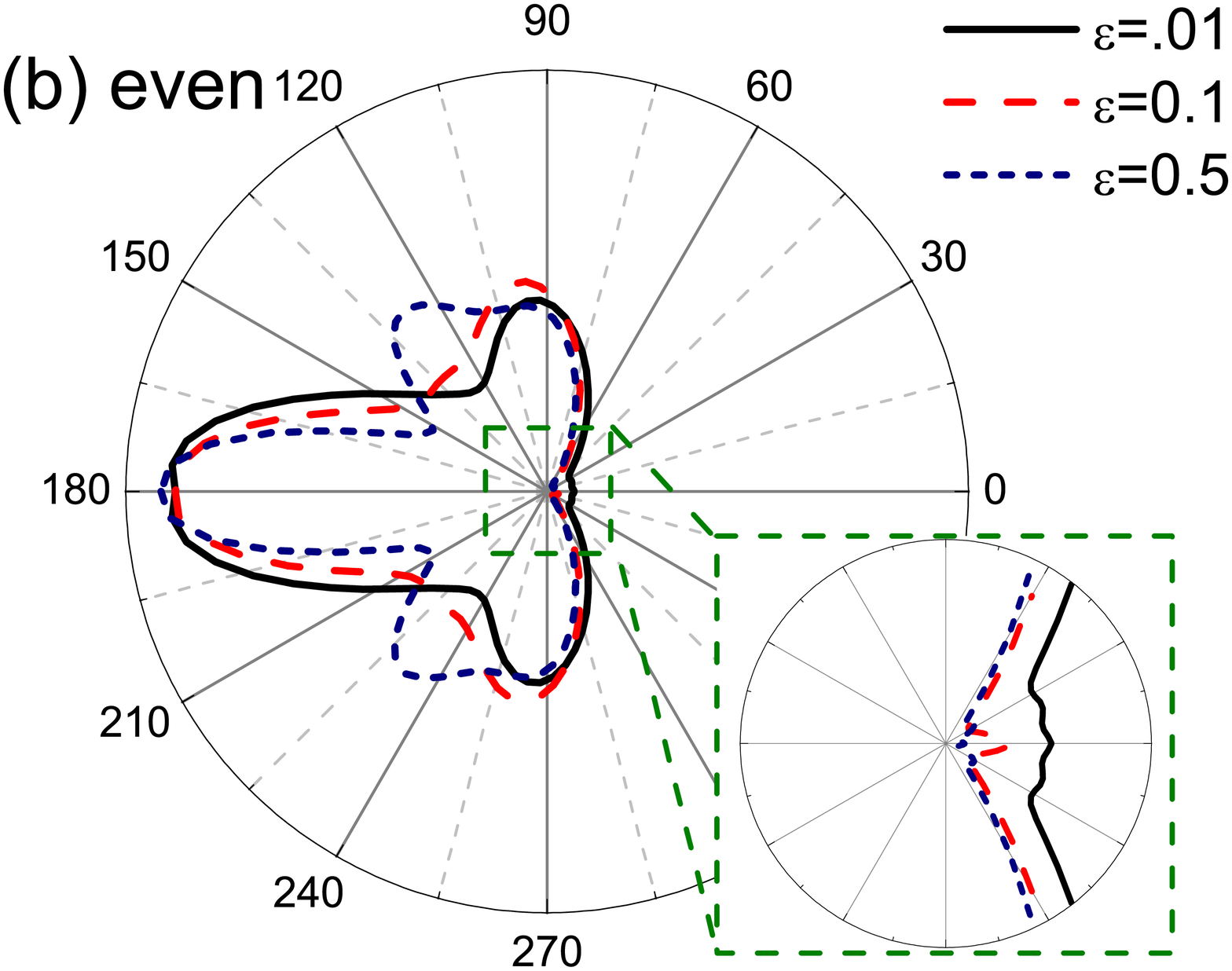}
\end{minipage}
\centering
\begin{minipage}[c]{0.7\linewidth}
\includegraphics[width=.99\columnwidth]{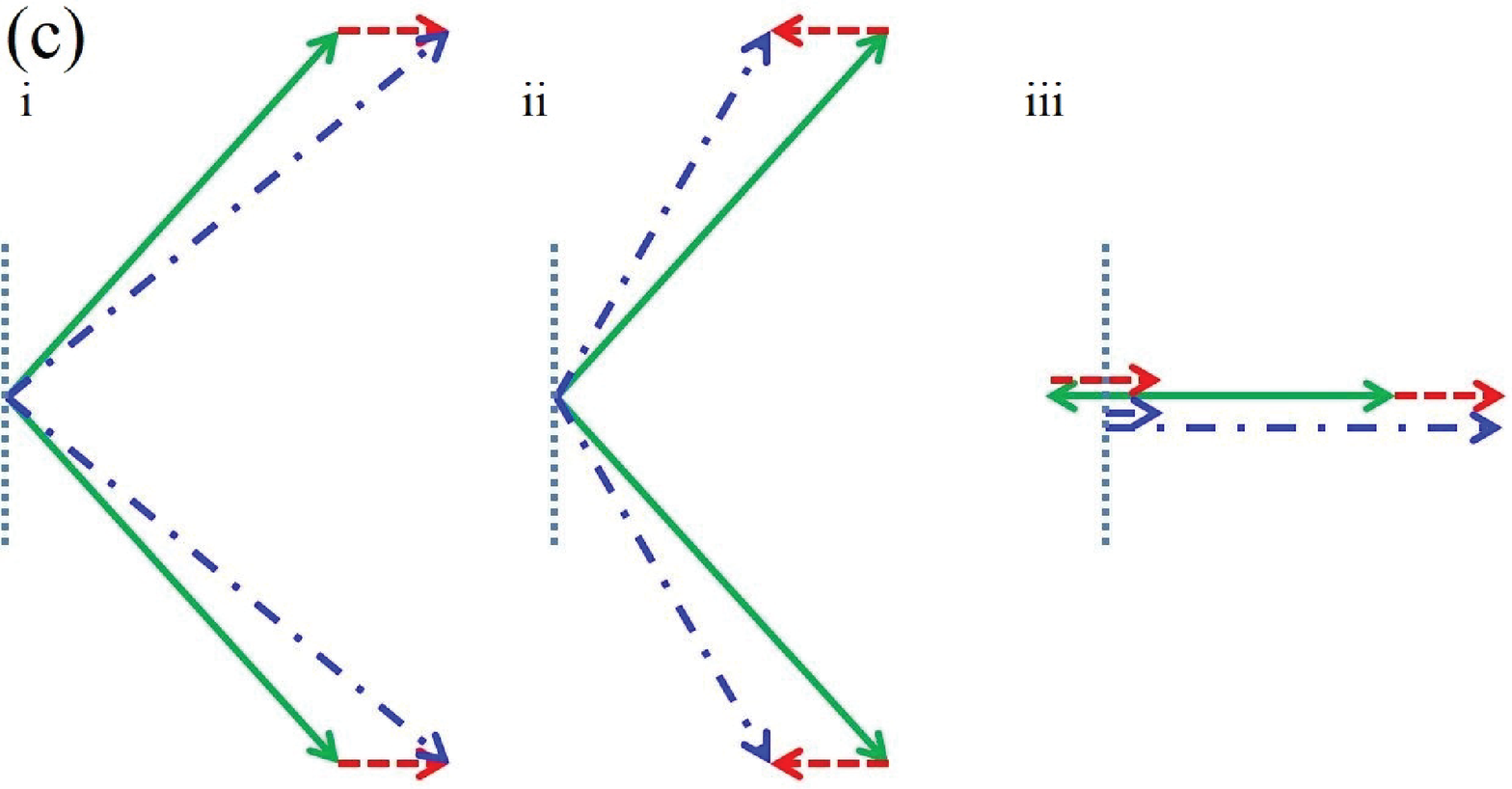}
\end{minipage}
\caption{Angular distributions versus the angle difference $\theta_{12}$ between two photoelectrons for laser assisted single XUV photon DI for (a) equal energy sharing $\epsilon=0.5$ and (b)extremely unequal energy sharing $\epsilon=0.01$.
The solid black lines are the MADs for DI with XUV laser field only, the blue dot lines show the MADs for the DI of electrons absorbing odd numbers of photons, the green dashed-dotted line shows the case for electrons absorbing even numbers of photons, and the red dashed lines gives the results containing both the odd-parity and even-parity electrons.
}\label{fig:mad}
\end{figure}

We show the mutual angular distribution for odd parity in Figs.~\ref{fig:mad}(a) and even parity in the Figs.~\ref{fig:mad}(b) respectively. 
The detail discussion of Figs.~\ref{fig:mad}(a,b) can be found in the reference \cite{Liu2014}, and we will skip it here.
To understanding the splitting of MADs, and enhancement/enabling of side-by-side/back-to-back emission pattern of photoelectrons in the Figs.~\ref{fig:mad}(a,b), in the Figs.~\ref{fig:mad}(c) we schematically display two typical modification to the momentum by adding an IR-laser field to the single-XUV-photon ionization at (i,ii): equal energy sharing and (iii): extremely unequal energy sharing. In the case of weak laser field, the Up of IR field is small, and DI caused by IR field is negligible. In this case, we can treat the IR field classically as the perturbation after the photoelectrons absorb a single photon from the XUV pulse.
The green solid lines and arrows stand for the initial momentum of the two emitted photoelectrons by absorbing the XUV photon only. 
Figs.~\ref{fig:mad}(c)(i,ii) displays how the weak controlling laser modify the photoelectrons shift to
parallel/anti-parallel emissions, respectively. The DI photoelectrons gain initial energy by
absorbing one XUV photon, and propagate in the laser field. The laser field can streak
the photoelectrons, and these electrons can obtain equal amounts of extra energies and
momenta (indicated by red dashed lines and arrows). These momenta can be parallel (i) or anti-parallel (ii) to the electric field of the IR laser pulse. In the case (i), initial and extra momenta both point to the left, the angle difference of their total momenta (plotted as blue dotted-dashed lines and arrows) therefore decreases, which means the photoemission shifts to side-by-side emission. On the other hand, panel (ii) displays the photoemission shifting to back-to-back emission. While in the panel (iii) depicts how the laser can enable side-by-side emission. The initial momenta of the two electrons are asymmetrical. One electron takes almost all the energy, while the other takes little. Because of correlation, the two photoelectrons go in opposite directions. When the IR laser is added, both electrons obtain extra momenta, as shown in panel (iii), both pointing to right. Since the extra momentum is greater than the smaller initial momentum, the momentum of this electron will therefore be turned 180$^{\circ}$, the same as the other electron. Since back-to-back emission is relatively strong only in the extremely unequal energy sharing cases, side-by-side emission is also observable for small energy sharing when the IR laser is presented, although it is very small (see Fig.~\ref{fig:mad}(b)). 

\section{Conclusion}\label{sec:Con}
In conclusion, the DI process of He caused by moderate strong XUV radiation and a weak IR laser field was investigated. It has been shown that the probability densities for the odd-parity photoelectron and even-parity photoelectron in joint energy distribution show opposite changes with the increasing intensity of the assisted-IR laser field, and the results in the joint angular distribution shows consistent agreement with it. With adding a weak assistance laser field, the competition between odd and even parity photoelectrons is raised and become stronger and stronger with laser intensity increasing within the weak limit. Moreover, we studied the dependence of joint angular distribution on the energy sharing with and without the assisted-IR laser field. The results indicates that the components of emission pattern for the photoelectrons changed by adding an assisting-IR laser field in the DI process produced with XUV pulses . Finally, we show and demonstrate how the modification to mutual angular distributions for equal energy sharing and extremely unequal energy sharing, and find that the enhancement of back-to-back emission in the case of assisted-IR laser field presented is referred to the odd-parity photoelectrons.

\section*{Acknowledgements}
This work is supported by the National Natural Science Foundation of
China under Grant Numbers 11774131, 91850114.

\addcontentsline{toc}{chapter}{References}

\bibliographystyle{iopart-num}
\bibliography{kitebib01}

\end{document}